# UDP: an integral management system of embedded scripts implemented into the IMaX instrument of the Sunrise mission

Rafael Morales Muñoz[a], Pablo Mellado[a], José Marco de la Rosa[b], and the IMaX Team
[a]Instituto de Astrofísica de Andalucía (CSIC), C/ Camino Bajo de Huétor, 50,
E-18080, Granada Spain.
[b]Instituto de Astrofísica de Canarias, C/ Vía Láctea, s/n, E-38205, La Laguna (Tenerife) Spain.

## ABSTRACT

The UDP (User Defined Program) system is a scripting framework for controlling and extending instrumentation software. It has been specially designed for air- and space-borne instruments with flexibility, error control, reuse, automation, traceability and ease of development as its main objectives. All the system applications are connected through a database containing the valid script commands including descriptive information and source code. The system can be adapted to different projects without changes in the framework tools, thus achieving great level of flexibility and reusability. The UDP system comprises: an embedded system for the execution of scripts by the instrument software; automatic tools for aiding in the creation, modification, documentation and tracing of new scripting language commands; and interfaces for the creation of scripts and execution control.

## 1. INTRODUCTION

Software (SW) for air- and space-borne instruments is often specifically designed for the project. This is not strange given the specificity of the various projects. Unfortunately, this peculiarity makes the SW not easily reusable in different instances. There are however some issues that are common to this kind of instruments. Thus, exploiting similarities seems advisable mostly when a given team is likely to engage in several developments. In this contribution, we report on the design of an integral scripting system that addresses some of these common issues, aiming at being as reusable as possible..

A real project has been used as the first working scenario, namely, the IMaX (Imaging Magnetograph eXperiment) [11],[1],[8] solar magnetograph, an instrument for the *Sunrise* [14],[12],[15] mission..

The present document is divided into five sections. Section 2 introduces the IMaX instrument and the *Sunrise* mission. Section 3 introduces the working scenario and the main features of the problem to be solved. Section 4 explains the developed system in detail. Section 5 comments on the application of the system in other real instruments. Finally, Section 6 presents the final conclusions and some proposals for future work.

## 2. IMAX AND SUNRISE

*Sunrise* is a balloon-borne mission including a solar telescope with an aperture of 1 m, a project developed in co-operation among institutes in Germany, Spain, and the USA. The *Sunrise* telescope feeds two focal-plane instruments, namely, the *Sunrise* Filter Imager (SUFI), which takes images in four wavelength bands in the blue and the UV with unprecedented spatial resolution; and IMaX which is an experimental solar magnetograph that produces very high resolution vector magnetograms of the solar surface. The first long-duration flight of *Sunrise* over the Arctic is planned in the summer of 2009.

As in any other similar instrument, a clear separation between the flying SW/hardware (HW) and the ground SW/HW components is mandatory. The two components should communicate through some communication channel when available. Following usual conventions, we shall divide the instrument software into two main blocks: the Flight SW (FSW) and the Electrical Ground Support Equipment (EGSE) SW, each running on a different computer. The IMaX embedded computer is based on an Intel Pentium 3 Mobile microprocessor at 1.6 GHz with 512 MB RAM, 4GB of flash disk massive storage and one PC/104-Plus slot used by the DSP (Digital Signal Processor)-FPGA (Field Programmable Gate Array) board. The HW selected to perform the digital signal processing tasks is two Texas Instrument Modules (TIM) 374-6713. Each TIM has two 6713 DSPs running at 225 MHz with 256 MB SDRAM and one Virtex2-2000 FPGA Each DSP has 8 KB of Level 1 cache and 256 KB of Level 2 cache. The EGSE computer HW is a general purpose PC.

The IMaX FSW is made up of three main functional blocks: the Control SW (CSW), the DSP SW (DSW) and the Embedded Script Runner (ESR). The CSW coordinates the operation of the other two blocks and communicates with the main *Sunrise* computer. The main CSW activities are image compression, packing, telemetry (TM) and telecommand (TC) management, and subsystem control. The main DSW activities are image acquisition and processing (other than compression), and low-level control of the various actuators. The ESR is in charge of validating and executing scripts as described in the following sections.

The EGSE SW consists of the communication modules with the instrument (EGSE-COM), the graphical-user interface for interacting with the instrument, and the implementation of the proposed User Defined Program (UDP) system. The interaction with the instrument includes reception and monitoring of TM data (about the status of the instrument and/or the science) and sending of TCs for commanding the instrument.

The FSW and the EGSE SW communicate through two communication links: a 600 KB/s bandwidth and another 5 bit/s bandwidth. The use of the higher bandwidth link is limited to situations where there is a direct line of sight to the instrument. The low bandwidth link will be in use for more than 90% of the flight.

## 3. WORKING SCENARIO

As explained in the previous section, bandwidth availability is one of the main problems of our working scenario. An autonomously scheduled working plan (SWP) is thus necessary because of the possible temporary losses of communication. The SWP will be stored onboard and can be changed online if needed and communications are on. For IMaX, the SWP is actually an Observation Plan (OP) which is a scheduled set of tasks related to the astronomical observations.

An important point is how to update the OP during flight. A first solution is reached by including the OP within the FSW code. This implies that every time a modification in the OP is done, the whole, newly compiled FSW version should be transmitted to the flying instrument with a prohibitively low bandwidth. A second solution is patching the FSW executable. This solution requires the development of some extra tools for performing the patching and depends on how the compiler generates the code. Sometimes, the change of a single line of code can lead to hundredths of bytes in patching. Patching is a hard task and produces source code that is difficult to maintain as many dependencies must be monitored. A third solution is scripting.

### 3.1 Scripting

The OP is made up of a sequence of SW functions (commands). Examples of usual commands are: setting observing mode parameters, starting an observation, stopping the current observation, performing calibration tasks, etc. The idea is keeping the OP in a file (script file) storing the command sequence to be executed, so that changing the script, the OP is updated. An embedded script runner (ESR) in the instrument FSW is needed to perform this task. This ESR should comply with the following characteristics as much as possible:

i) Simplicity, both in code complexity and size. Balloon- and space-borne instruments are likely to include hardware with limited resources. Hence, extensive computing is not advisable.. The script language for building OPs should be simple, therefore precluding the use of a virtual machine like Javascript or Microsoft's .NET).

ii) Security, forcing the removal of potentially dangerous programming features. This can easily be achieved when you create your own domain specific language [10], as is the case.

iii) Reliability, usually implying the availability of the source code or an extensive validation campaign for a third party solution.

iv) Flexibility, given the ease of access to the SW functionality from the script.

v) Scalability, allowing an easy addition and maintenance of the SW functionality accessible from the script.

vi) Compatibility with your instrument FSW programming language (ANSI C++ in IMaX's case).

Since each script command is executed within the FSW, they can be used not only for the execution of an OP but also for all the instrument low-level functionalities programmed in the FSW (characteristics iv and v). Allowing the access to all this functionalities provides great help in integration, testing and validation stages while not increasing the code complexity, as far as all the functionalities are already available in the FSW. The user will therefore be able to create scripts with different granularities ranging from very low-level functionality (integration and testing) to high level (OP). These different levels of programming can also be imposed depending on authorization policies just by restricting access to different commands depending on the user's role (engineer, scientist, etc.).

Some products (both commercial and free) have been analyzed and finally the decision has been made that, in order to fulfil the previously mentioned characteristics, this SW development will start from scratch. A list of the analyzed products can be found in references [4],[7],[13],[16]. The available products do not adequately comply with, at least, characteristics i.

## 4. DEVELOPED SYSTEM DESCRIPTION

The main objective of the developed system is automating as much as possible (always pursuing the characteristics mentioned in Section 3.1) the processes of:

- script command creation, including source code generation for the ESR;
- script execution control, including script interpretation and communication with the ESR embedded in the FSW;
- scripting itself, through the appropriate interfaces;
- documentation for all the automatic output.

In our system, each command is called a DECO (DEfined COmmand) and each script is called a UDP (User Defined Program).

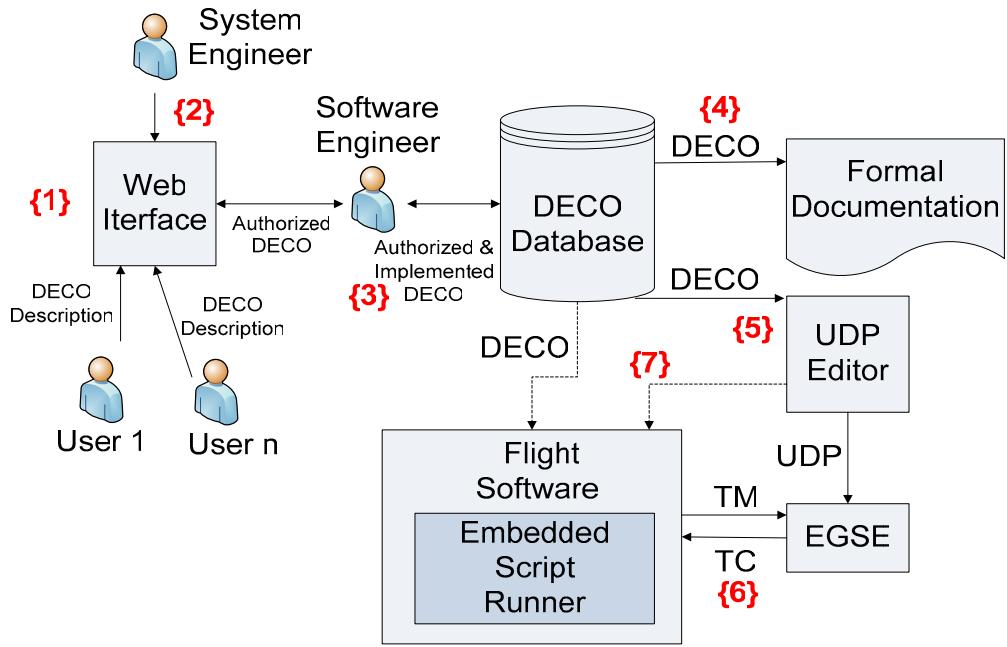

Figure 1. General Working Scheme.

By using the Web Interface {1}, the users (engineers and scientists, typically) describe (DECO Description) the desired DECO in a detailed way (name, description, parameters, parameter limits, etc). This description is sent to the System Engineer, who must check and authorize the DECO before storing it in the DECO database (DDB) {2}. These checks must include semantic validation (the DECO must perform a reasonable action), parameter limit validation and redundancy validation (there is no other DECO that implements the same functionality). Once the DECO is accepted, the SW Engineer implements it by using all the input information and stores it in the DDB {3}.

The DDB also stores all the changes performed to an authorized DECO, allowing the review of its history and evolution. Formal documentation, including the DECO catalogue and documentation for the DECO's source code is automatically generated from the DDB {4}. The same DB is shared by the Web Interface (with read and write rights), the UDP Editor (with read rights) and the instrument FSW (with read rights), thus maintaining the unity and consistency of the DECO definition.

The UDP Editor gets the DECO definition from the DDB {5} and allows building UDPs that will be transferred to the EGSE which will compile it to a binary format and transmit it to the FSW encapsulated in a TC {6}, as explained in Section 4.1

The system includes a template-based, code generation feature that creates FSW compatible source code (C++ in the IMaX case) for all the DECOs in the database. The resulting source code is built as a combination of the mentioned templates and the DECO implementation code that the SW Engineer inserted in the DDB {3}. This automation eliminates the bugs derived from the common code included in the templates, thus reducing the coding effort significantly. The documentation associated to this code is also generated automatically.

Special emphasis has been done in tracking all the DECO life cycle: creation, modification and deletion, so any DECO edition implies a DDB update. Every DECO is tagged with a version which is incremented with any DECO edition. This DECO feature is used to implement a version-based validation mechanism that avoids outdated DECOs to be executed by the ESR. This feature, together with System Engineer revision, constitutes the two main system safety mechanisms.

The system has been developed using open source tools in order to reduce budget and to guarantee an easy translation to a concrete platform (HW and SW): Java (UDP Editor), Perl, JavaScript, Ajax (Web Interface), PHP, MySQL (for the DDB) and C++ compilers (FSW). The system has also been designed to be instrument independent, to allow working with multiple and distributed users (Web Interface), and to allow centralized storage and supervision (DDB). The described system layout makes it easy for a user to request lacking functionality and also for the engineers to implement and document it, thus reducing significantly the time required for the software extension task.

The developed system includes the possibility of being deployed completely (except for the FSW) in a single computer. This configuration is especially useful for very restrictive environments (e.g. militarized zones, integration places with no network access).

The following sections describe the UDP execution chain, the DECO building, the UDP creation and the DDB in detail.

### 4.1 UDP execution chain

Basically five main SW blocks from the architecture presented in Figure 1 are related to UDP creation and execution: Web Interface (Section 4.2), UDP Editor (Section 4.3), EGSE-COM, FSW and ESR. To better understand the following sections, a more precise explanation of the UDP chain of transformations and communication events from its creation in the UDP Editor to its actual execution by the ESR is introduced.

The UDP execution chain can be divided into three main steps:

i) UDP construction. Two interfaces provide support for UDP creation. The first one is the UDP Editor, that allows the creation of the DECO sequence and the second one is the Web Interface that allows requesting new DECOs, in case that some functionality is missing in the current DECO list.

ii) UDP compilation and transmission. The FSW is linked to the EGSE by a (typically low-bandwidth) communication channel so that UDPs are compiled by the UDP Editor to a compact and compressed byte sequence, hence optimizing the allowed bandwidth usage. This compiled UDP is forwarded to the EGSE-COM which transmits it to the FSW encapsulated in a TC.

iii) UDP execution. The FSW receives the UDP-encapsulating TC, extracts the compiled UDP from the TC and forwards it to the ESR. The ESR decompresses the UDP and checks, with the local copy of the DDB, that all the DECOs found in the UDP sequence are valid. Parameter limits and DECO versions (in order not to execute a different version of the DECO) are checked. If no error is found, the UDP sequential execution starts; otherwise, an error is reported back to the EGSE in the form of a TM.

Note that in a flight scenario, where the FSW is on board, the only visible feedback that UDPs can generate at execution time are TMs sent back to the EGSE.

The TM/TC communication infrastructure which supports the UDP execution chain is common to all air- and space-borne instruments, so the implementation of that execution chain in other such instruments should be easy.

### 4.2 DECO building: Web Interface

The system users are able to request the implementation of new DECOs through a Web interface. This way they can specify the details of the DECO including the proposed name for the DECO, a description of its functionality and the associated parameters.

A user needs a username and a password to log into the system. Every user is able to create its own labels and to assign them to different DECOs, for arranging purposes. Each label can be associated or not to one user. At the same time, a DECO may be linked to more than one label, even from different users.

The labels have internal relationships, where a label may be parent of several others. Then, a tree structure of labels is normally obtained.

Figure 2. Web Interface.

The Web interface offers an access to the system through the Internet from any computer with a Web browser and allows the users to concurrently log to a centralized online database.

Once the DECO is defined using the Web interface, and confirmed by the System Engineer, a SW Engineer has to write the source that implements the requested functionality. Therefore, new source code is added to the instrument SW. Inside this source code, the DECO can have complex structures (conditionals, variables, etc.) not available in the UDP language.

**4.3 UDP creation: UDP Editor**

The UDP and DECO infrastructure allows the programmer to develop generic UDP editors.

The UDP language syntax is extremely simple, basically for two reasons: in order to keep a small and lightweight ESR and because the studied OPs do not need complex computation structures (see i) and ii) in Section 3.1). The UDP language does not allow conditional constructions or variables. The only program flow control structure is a fixed number of iterations loop. As stated in the previous section, more complex control structures are available at DECO level.

If the available DECOs do not allow the user to program a particular operation, the new functionality can be requested through the use of the Web Interface (Section 4.2), following the process described in Section 4. Given the centralized nature of the DDB, once the new functionality has been added to the FSW, the corresponding DECO will be immediately available for the user in the UDP Editor.

The first UDP editor developed for the UDP system looks as shown in Figure 3.

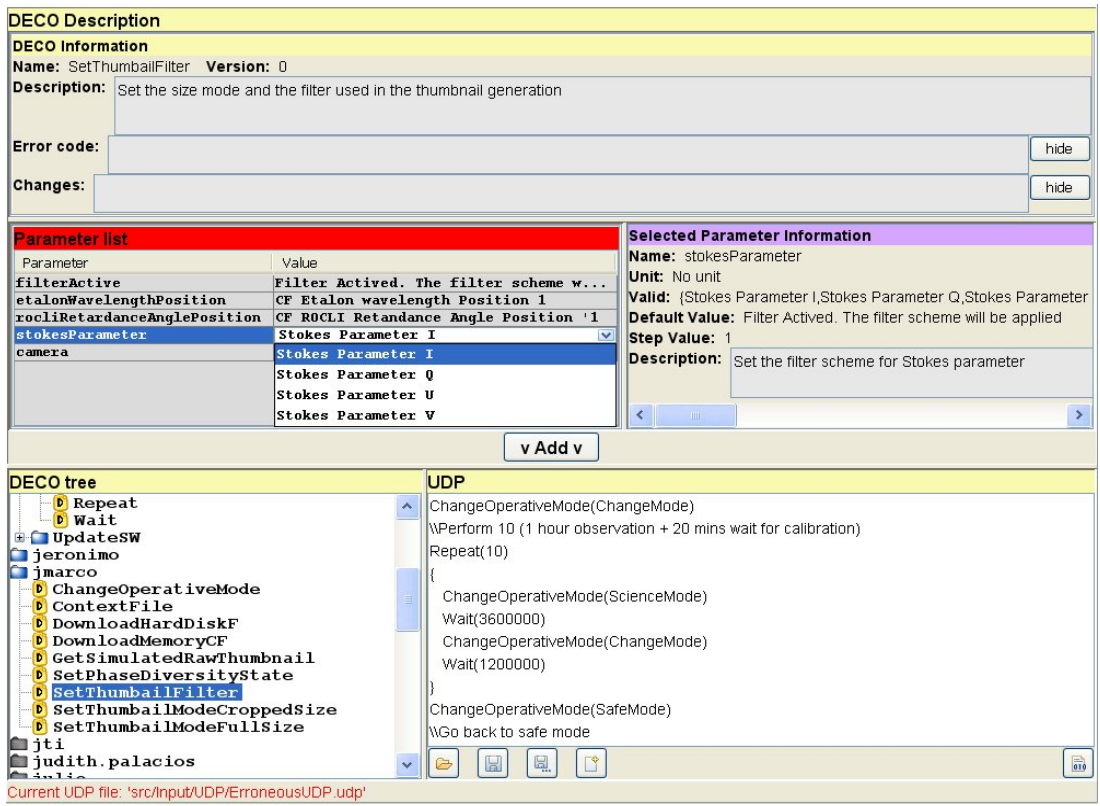

Figure 3. UDP Editor.

The UDP Editor is divided into four main panels: selected DECO information (up), selected parameter information (middle), DECO tree (down left) and UDP panel (down right). Their functionalities are as follows:

- The selected DECO information panel shows all the information regarding the currently selected DECO, which can be selected in the DECO tree or in the UDP panel.

- The selected parameter panel shows a list of all the parameters associated to the currently selected DECO and the information associated to the currently selected parameter. The edition of the value for the selected parameter is allowed in order to define the parameter values for insertion of new DECOs in the current UDP or for the modification of previously inserted DECOs.

- The DECO tree allows the selection of a DECO for insertion. It also shows the associations among DECOs and labels as defined in the DDB. Some authorization mechanisms for masking out DECOs (e.g. integration DECOs) can be implemented based on this tree structure, given labels can be associated to users.

- The UDP panel shows the currently built UDP. A simple UDP which can perform a series of observations for a given instrument is shown in Figure 3. The UDP panel allows the selection of a DECO for parameter tuning, duplication, or deletion. Common edition features as changing the DECO's order or UDP file saving/loading are also implemented.

The main advantage provided by the UDP system regarding UDP edition is that all the information relative to DECOs is taken from the centralized DDB, thus maintaining the coherence with the rest of the system. Adding or copying DECOs into a UDP is done through mouse clicks, completely eliminating the possibility of syntactic errors. Besides, input parameter value ranges are also checked against the information stored in the central DDB, so the insertion of values out of range is not possible.

### 4.4 Database layout

The DDB is stored by a relational database management system (RDBMS [5],[6]). Using the Entity-Relationship [2] (ER) paradigm, the structure of the current database is shown in Figure 4.

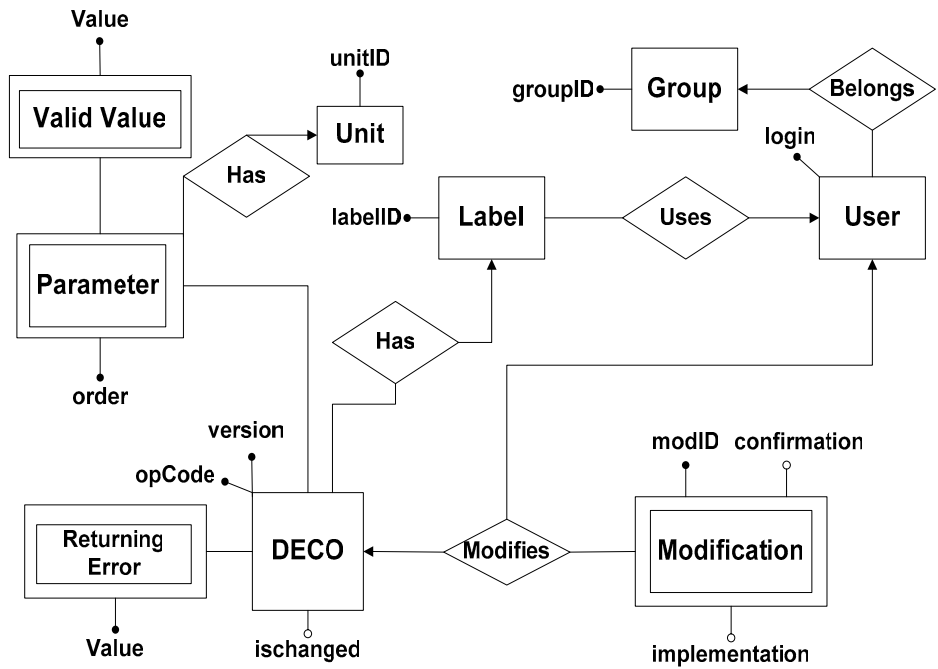

Fig. 4. DECO Database (DDB).

Only the most essential attributes are shown in the figure. All primary keys and some important attributes are present, though.

The main tables of the database are *DECO* and *Parameter*, which describe all the information about DECOs needed. Tables *User* and *Group* are related to user login, *Modification* stores DECO changes for tracing purposes and *Label* stores the DECO arranging labels.

An important disadvantage of relational database management systems (RDBMSs) is that they are not usually suitable for an air- or space-borne instrument FSW. CPU and memory restrictions of the HW associated with this kind of instruments make this solution unfeasible. The use of an XML file generated by the DBMS containing only the information needed by the instrument has been the best solution found. The XML support in the FSW is less resource consuming than the RDBMS. This XML file is generated on request by the Web server and used by the FSW and EGSE SW. Anyway, a plain text custom format or any other standard format could also be used (e.g. JavaScript Object Notation JSON [9]) just by changing the RDBMS output.

# 5. APPLICATION IN INSTRUMENTS

## 5.1 IMaX

As stated in **Error! Reference source not found.**, IMaX is the first instrument to use the UDP system. The objective of supporting the management of the OP was fulfilled and besides, given the ease of access to all the SW functions, the system helped us greatly during the instrument integration stage. In this stage, SW and HW are running and an amount of SW modification and test cycles is usually required in order to reach the desired functionality. Some of these tests consist in checking some HW or SW parts in very low detail and involve the generation of a big amount of code that will be removed in the final SW version. Note that, during this stage, the development team is under big pressure and this extra code is usually generated without a strict control. The proposed system organizes traces and stores the history of all these modifications.

Additionally, the UDP system will be extensively used during the instrument SW validation stage, building specific SW, in the form of DECOs, to test specific SW functionality.

## 5.2 Medusa

Medusa is an instrument in development for the ESA's space mission Exomars [17]. The UDP system is currently in use in the first prototype of the instrument to check the HW performances and capabilities.

## 5.3 Adaptation to other instruments

In order to adapt the UDP system to other architectures, the construction of a new DDB (using the Web Interface) and the implementation of the ESR are the only requirements. The front-end (Web Interface and UDP Editor) can be kept as is. The ESR implementation is simple, both algorithmically and in the data structures used, so it can be easily exported to more restrictive languages as the ones used in microcontrollers.

# 6. CONCLUSIONS AND FUTURE WORK

In the present paper an integral scripting system for air- and space-borne instruments has been presented. The system constitutes a solution for the bandwidth and need for autonomy problems inherent to that kind of instruments. This system is easily adaptable to new instruments, thus achieving a good degree of reusability. The current application of the system to two different real projects is a trial of it.

Thanks to all the provided automation, the proposed system has aided notably in the SW development. It has also shown really helpful in integration and validation stages. The access to low-level FSW functionalities is necessary in both stages and the finding of some functionality lack is not unusual during integration. Access to low-level functionality has been trivial with the given implementation and the addition of new functionalities has been quick and not error prone. Only small pieces of code have to be developed by the SW Engineer since much of the new source code has been generated automatically. Documentation for the newly developed functionality is also automatically generated.

The centralized management of information that is shared by different and independently running SW applications (FSW, EGSE-SW) has greatly enhanced the coherence of the system and made its maintenance much easier.

The Web Interface allows the formal coordination of different users (engineers or scientists) asking for instrument functionality at different levels. It also guides all the functionality implementation process. These tasks are achieved with very little overhead and with an accurate trace of each request.

As a safety mechanism, the proposed system performs several checks on UDPs. Syntactic and semantic checks are performed in all DECO definition, DECO implementation, UDP construction and UDP execution phases.

The currently implemented UDP system is very functional, although some interesting improvements could be added. Some guidelines for future work are:

- DECO access control: As described in Section 4.2, DECOs are labelled. By extending that label structure, a user/group based authorization mechanism can be implemented.
- Currently, all DECO source code is generated from the DDB and compiled and linked with the FSW in a static way. The implementation of some way of dynamic linking would reduce the amount of data to upload in case the DECO functionality should be updated.

- The main project that has led the development of the UDP system does not require a complex flow-controlling scripting language. This scenario can change for different instruments so the development of a more expressive UDP language can be studied. For this task, Domain Specific Languages' (DSLs) techniques [10][3] can be applied.

## ACKNOWLEDGEMENTS

This work has been funded by the Spanish Ministerio de Educación y Ciencia through project ESP2006-13030-C06-02, partially using European FEDER funds.